\documentclass[preprint,showpacs,preprintnumbers,amsmath,amssymb, showkeys,superscriptaddress,titlepage]{revtex4-2}
\usepackage{graphicx}
\usepackage{dcolumn}
\usepackage{bm}
\usepackage{hyperref}
\usepackage{textcomp}

\begin{document}

\title{An enhancement of formation of unstable $^{8}$Be nucleus with the growth of $\alpha$-particle multiplicity in fragmentation of relativistic nuclei}

\author{A.A. Zaitsev}
\email{zaicev@jinr.ru}
\affiliation{Joint Institute for Nuclear Research (JINR), Dubna, Russia}
\affiliation{Lebedev Physical Institute, Russian Academy of science, Moscow, Russia}
\author{P.I. Zarubin}
\affiliation{Joint Institute for Nuclear Research (JINR), Dubna, Russia}
\affiliation{Lebedev Physical Institute, Russian Academy of science, Moscow, Russia}
\author{N.G. Peresadko}
\affiliation{Lebedev Physical Institute, Russian Academy of science, Moscow, Russia}

\begin{abstract}
In this paper, the correlation between the formation of the unstable $^{8}$Be nucleus and accompanying $\alpha$ particles in the fragmentation of relativistic $^{16}$O, $^{22}$Ne, $^{28}$Si, and $^{197}$Au nuclei in a nuclear track emulsion is investigated. The $^{8}$Be 
decays are identified in a wide energy range by invariant masses calculated from 2$\alpha$-pair opening angles. The adopted approximations are verified by data on fragmentation of $^{16}$O nuclei in a hydrogen bubble chamber in a magnetic field. An increase in the $^{8}$Be contribution to the dissociation with the growth of $\alpha$-particle multiplicity is found.
\end{abstract}

\maketitle 

\section{Introduction}
The phenomenon of a multiple fragmentation of relativistic nuclei has a hidden potential for the study of nonrelativistic ensembles of H and He nuclei \cite{1}.
The decays of unstable nuclei $^{8}$Be $\to$ 2$\alpha$ and $^{9}$B $\to$ 2$\alpha p$
and the Hoyle states HS $\to$ 3$\alpha$ \cite{2} are of actual interest.
Each of these unstable states has extremely low decay energy. Consequently, against the background of other relativistic fragments, they should appear as pairs and triplets with the smallest opening angles. The lifetimes of $^{8}$Be (5.6 eV) and $^{9}$B (540 eV) unstable states and HS (9.3 eV), which are inversely proportional to their widths, make them full-fledged participants of relativistic fragmentation. Their decay products
are formed at range from several thousand ($^{8}$Be and HS) to several tens ($^{9}$B) of atomic sizes, i.e., over times that are many orders of magnitude longer than the times of generation of other fragments. The predicted sizes of these states are exotically large \cite{3}. All these facts make HS, $^{9}$B, and $^{8}$Be extremely interesting
objects for understanding the microscopic pattern of fragmentation and signatures in the search for more complex nuclear–molecular structure states decaying through them.

The identification of decays requires recovering invariant masses: $Q_{2\alpha}$ of 2$\alpha$ pairs, $Q_{2\alpha p}$ of 2$\alpha p$ triplets, and $Q_{3\alpha}$ of 3$\alpha$ triplets, respectively. Generally, invariant mass $Q = M^* – M$ is set by the sum $M^{*2}$ = $\Sigma(P_i\cdot P_k)$, where $P_{i,k}$ are the 4-momenta of the fragments and $M$ is their mass. In the case of relativistic fragmentation, the application of this variable is feasible only within the nuclear track emulsion (NTE) method. The NTE layers with thicknesses from 200 to 500 $\mu$m, longitudinally exposed to the nuclei under study, provide accurate detection with 0.5 $\mu$m resolution of angles between the directions of relativistic fragments emitted in a cone $sin\theta_{fr}$ = $p_{fr}/P_0$, where $p_{fr}$ = 0.2 GeV/$c$ is the characteristic nucleon Fermi momentum in the projectile nucleus with momentum per nucleon $P_0$. In order to calculate $Q_{2\alpha}$ and $Q_{3\alpha}$, it is sufficient to assume that the fragments retain the momentum per nucleon of their primary nucleus and to use only their measured emission angles. Below, it is shown that, in the case of extremely narrow decays of $^{8}$Be and $^{9}$B, the measured contributions of $^{3}$He and $^{2}$H are small. Therefore, the He–$^{4}$He and H–$^{1}$H correspondence is assumed. Charges 1 and 2 are identified visually in NTE. The decay energies of these three states are considerably lower than the energies of the next excitation levels with the same nucleon composition, and the display of more complex excitations is unlikely for these nuclei. Therefore, a simple constraint on the invariant mass of the ensemble is sufficient for their identification. The sampling conditions tested in the most convenient cases of $^{9}$Be, $^{10}$B, $^{10}$C, $^{11}$C, and $^{12}$C isotope dissociation are $Q_{2\alpha}$($^{8}$Be) $\leq$ 0.2 MeV, $Q_{2\alpha p}$($^{9}$B) $\leq$ 0.5 MeV, and $Q_{3\alpha}$(HS) $\leq$ 0.7 MeV \cite{2}.

\begin{figure}
	\centerline{\includegraphics*[width=1.0\linewidth]{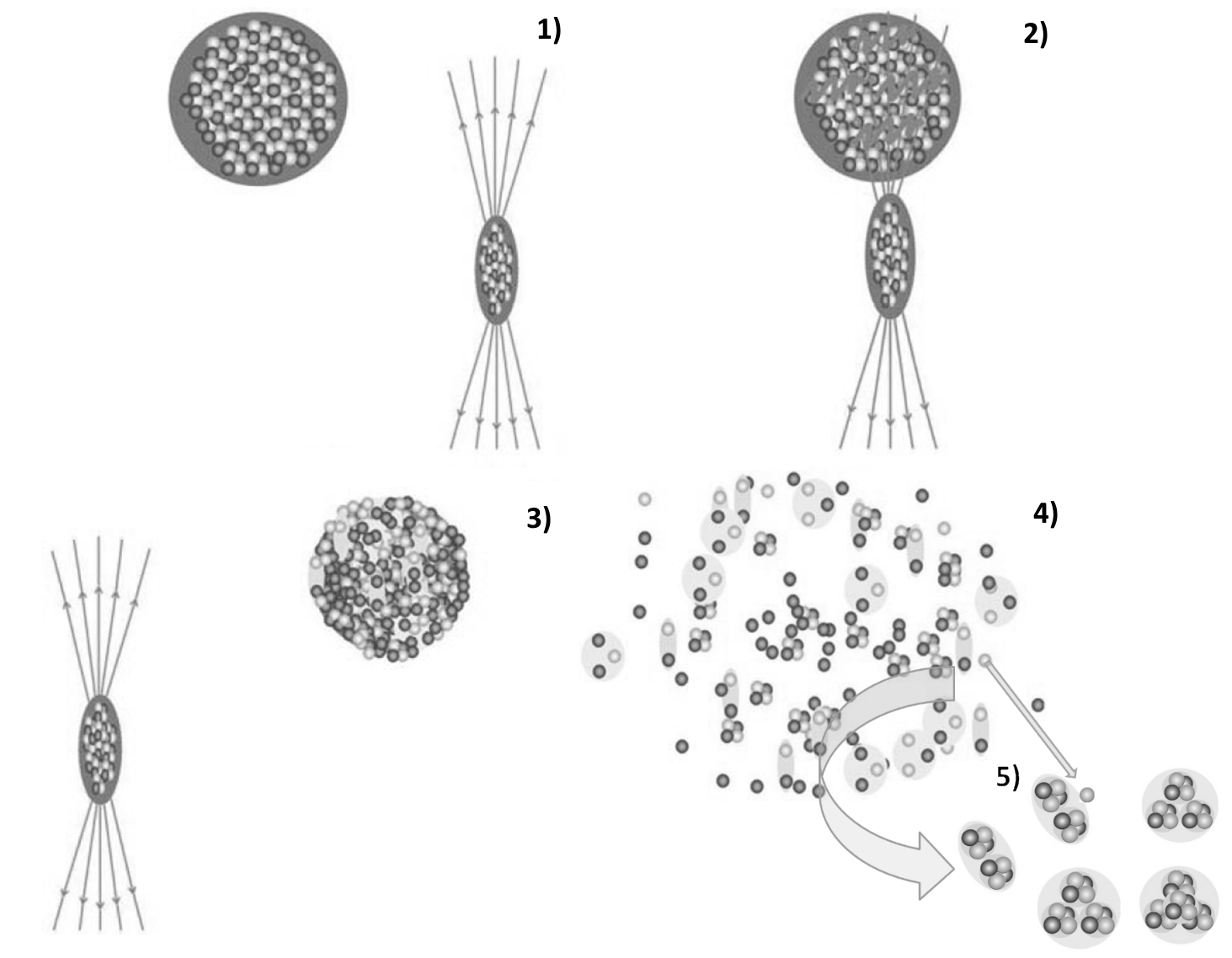}}
	\caption{Scenario of generation of multiple fragments: (1) approaching nuclei; (2) transfer of excitation to the studied nucleus; (3) transition to a system containing real lightest nuclei and nucleons; (4) decay of the system; and (5) capture and aggregation of some fragments into unstable states.}
\end{figure}

The unstable states are generated most efficiently during the coherent dissociation (or in ``white'' stars) that is not accompanied by the target fragments,because nucleon escape from the fragmentation cone is minimal in the events of this type. The analysis of ``white'' stars $^{12}$C → 3$\alpha$ and $^{16}$O $\to$ 4$\alpha$ showed that the fraction of events containing $^{8}$Be (HS) decays is \mbox{45 $\pm$ 4\%} (11 $\pm$ 3\%) for $^{12}$C and 62 $\pm$ 3\% (22 $\pm$ 2\%) for $^{16}$O \cite{4,5}. Supposedly, the growth of 2$\alpha$ and 3$\alpha$ combinations increases the contribution of $^{8}$Be and HS. This observation needs verification for heavier nuclei when the number of $\alpha$ combinations increases rapidly along with the mass number. The invariant mass method was used to estimate the contributions of $^{8}$Be, $^{9}$B, and HS decays to the relativistic fragmentation of Ne, Si, and Au \cite{2}. The features of the unstable state formation will be presented in this aspect.

It is possible that unstable states are present in the nucleus structure or emerge somehow at their periphery, which results in fragmentation. An alternative is that $^{8}$Be is generated in the interaction of generated $\alpha$ particles with the subsequent capture by accompanying $\alpha$ particles and nucleons and the emission of necessary $\gamma$ quanta or recoil particles. This may result in the increase in the yield of $^{8}$Be with the increasing multiplicity of $\alpha$ particles per event and possible $^{9}$B and HS decay through $^{8}$Be. Therefore, finding the relation between the formation of unstable states and the multiplicity of accompanying $\alpha$ particles is of particular interest. In Fig. 1, this scenario is presented in the reference system of a fragmenting nucleus: approaching nuclei, transfer of excitation, transition to a system containing the lightest nuclei and nucleons, its decay, and aggregation of some fragments into unstable states.

\section{INFLUENCE OF IDENTIFICATION OF H AND He FRAGMENTS}
The assumed approximations can be verified by data obtained upon irradiation with $^{16}$O nuclei with energy of 2.4 GeV/nucleon of a one-meter hydrogen bubble chamber VPK-100 placed in a magnetic field at the Joint Institute for Nuclear Research (JINR) \cite{6}. In this case, a peak is observed in the initial part of the opening angle distribution $\Theta_{2\alpha}$ of 2$\alpha$ pairs (Fig. 2), which corresponds to $^{8}$Be decays \cite{6}. When $Q_{2\alpha}$ is calculated using measured momenta $P_{\textrm{He}}$ of He fragments, which are recovered with an insufficient accuracy, the signal of $^{8}$Be almost disappears. It is still possible to record the momenta, as in the NTE case. The values of $P_{\textrm{He}}$ and $P_{\textrm{H}}$ normalized to the initial momentum $P_0$ (per nucleon) identify isotopes of He and H. According to Fig. 3, the condition $Q_{2\alpha}$($^{8}$Be) $\leq$ 0.2 MeV removes the $^{3}$He contribution, while the proton contribution is 90\% among the H fragments.

\begin{figure}
	\centerline{\includegraphics*[width=0.7\linewidth]{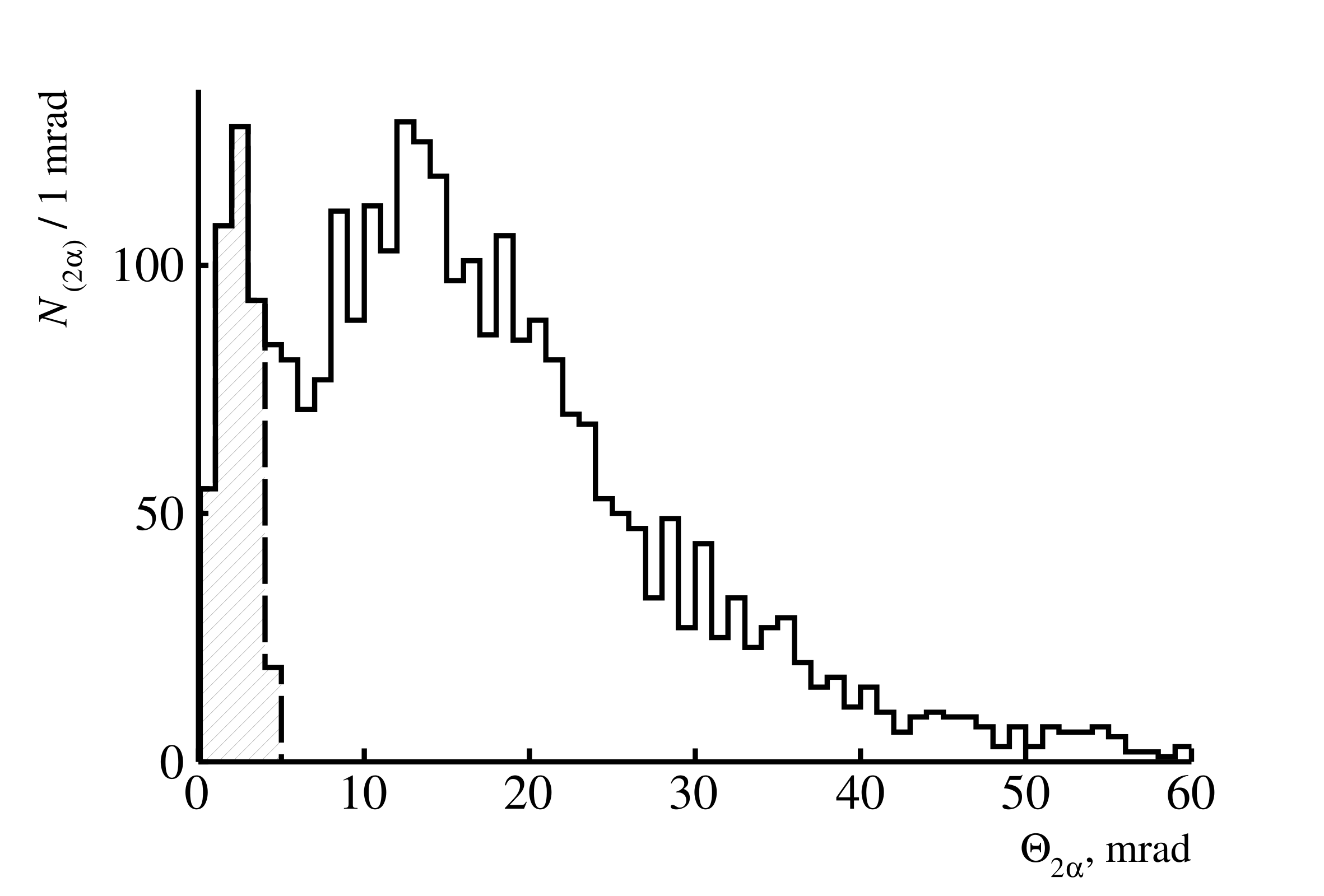}}
	\caption{Distribution of 2$\alpha$-pair combinations vs. opening
		angle $\Theta_{2\alpha}$ for the entire statistics (solid line) and under the
		condition $Q_{2\alpha}$($^{8}$Be) $\leq$ 0.2 MeV (dashed line) in fragmentation
		of $^{16}$O nuclei with momentum 3.25 GeV/$c$ per nucleon	on protons.}
\end{figure}

\begin{figure}
	\centerline{\includegraphics*[width=1.0\linewidth]{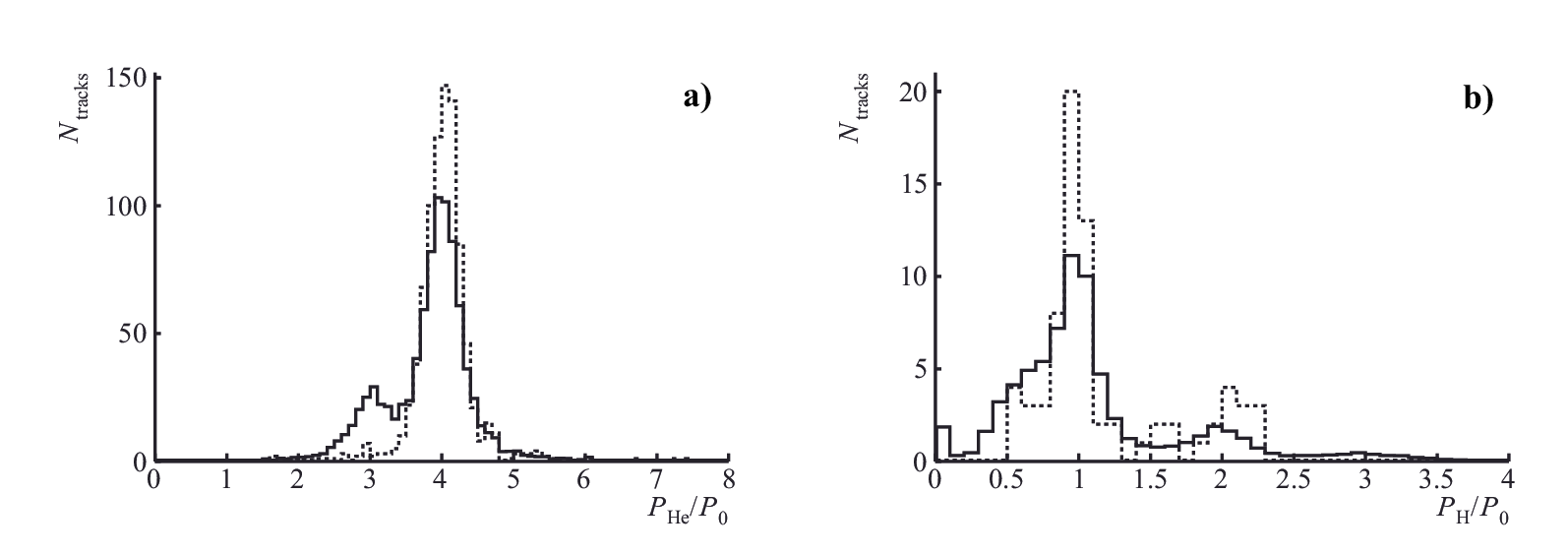}}
	\caption{Distribution of (a) H and (b) He relativistic fragments over the ratios of their measured $P_\textrm{H}$ and $P_\textrm{He}$ momenta to the initial
		momentum per nucleon $P_0$ (solid line) in fragmentation of $^{16}$O nuclei with momentum of 3.25 GeV/$c$ per nucleon on protons;
		samples with conditions $Q_{2\alpha}$($^{8}$Be) $\leq$ 0.2 MeV and $Q_{2\alpha p}$($^{9}$B) $\leq$ 0.5 MeV (dashed line) are indicated.}
\end{figure}

The distributions over invariant masses of all 2$\alpha$ pairs $Q_{2\alpha}$, 2$\alpha p$ triplets $Q_{2\alpha p}$, and 3$\alpha$ triplets $Q_{3\alpha}$ presented in Fig. 4 are calculated according to the angles found using the VPK-100. Distributions with selection
of $^{4}$He (3.5 $\leq$ $P_{\textrm{He}}$/$P_0$ $\leq$ 4.5), protons (0.5 $\leq$ $P_{\textrm{H}}$/$P_0$ $\leq$ 1.5), and $^{8}$Be ($Q_{2\alpha}$($^{8}$Be) $\leq$ 0.2 MeV) are added. The version with fixed momenta, which depends only on the fragment emission angles, displays $^{8}$Be and $^{9}$B peaks. A small number of HS candidates are present.

\begin{figure}
	\centerline{\includegraphics*[width=1.0\linewidth]{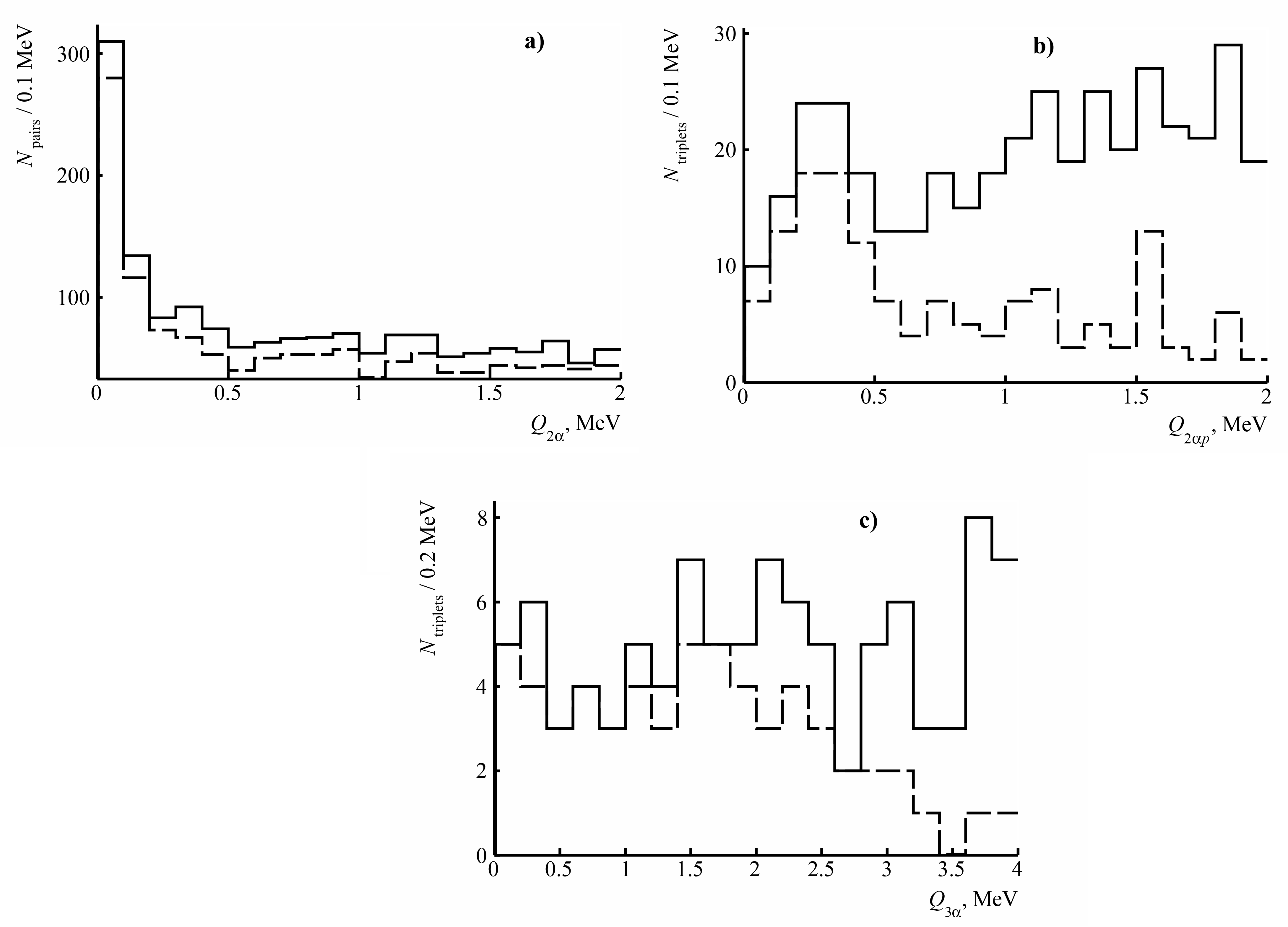}}
	\caption{Distribution of the events of fragmentation of $^{16}$O nuclei with momentum of 3.25 GeV/$c$ per nucleon on protons over
		(a) $Q_{2\alpha}$, (b) $Q_{2\alpha p}$, and (c) $Q_{3\alpha}$; distributions with conditions on $^{4}$He, protons, and $^{8}$Be 
		are added.}
\end{figure}

Table 1 shows the variations of unstable state contributions to events with multiplicity $n_\alpha$ of $\alpha$ particles (which are the identified $^{4}$He nuclei in this case). With the increase in $n_{\alpha}$, the probability of detecting $^{8}$Be increases. The increase in $n_\alpha$ leads to a relative decrease in $N_{n\alpha}$($^{9}$B), which can be explained by the decrease in the number of protons available to form $^{9}$B. On the contrary, $N_{n\alpha}$(HS) increases owing to the increase in the number of $\alpha$ particles available to generate HS. In the coherent dissociation $^{16}$O $\to$ 4$\alpha$, the fraction of HS decays with respect to $^{8}$Be is 35 $\pm$ 1\%, which does not contradict the value $n_\alpha$ = 4 in the harder interaction $^{16}$O + $p$ (Table 1). These facts indicate the universality of the $^{8}$Be and HS generation.

\begin{table}
	\centering
	\caption{Statistics of $^{16}$O + $p$ events in VPK-100 containing
		at least one $^{8}$Be decay candidate $N_{n\alpha}$($^{8}$Be), $^{9}$B, or HS provided
		$Q_{2\alpha}$($^{8}$Be) $\leq$ 0.2 MeV among $N_{n\alpha}$ events of fragmentation
		of $^{16}$O nuclei on protons with multiplicity $n_\alpha$}
	\begin{tabular}{ | c | c | c | c |}
		\hline
		$n_\alpha$ & 
		\begin{tabular}{c}
			$N_{n\alpha}$($^{8}$Be)/$N_{n\alpha}$ \\ (\% $N_{n\alpha}$)
		\end{tabular} & 
		\begin{tabular}{c}
			$N_{n\alpha}$($^{9}$B) \\ (\% $N_{n\alpha}$($^{8}$Be))
		\end{tabular} & 
		\begin{tabular}{c}
			$N_{n\alpha}$(HS) \\ (\% $N_{n\alpha}$($^{8}$Be))
		\end{tabular}   \\ \hline
		2 & 111/981 (11 $\pm$ 1) & 29 (26 $\pm$ 6) & — \\ \hline
		3 & 203/522 (39 $\pm$ 3) & 31 (15 $\pm$ 3) & 36 (18 $\pm$ 3) \\ \hline
		4 & 27/56 (48 $\pm$ 11) & — & 11 (41 $\pm$ 15) \\ \hline
	\end{tabular}
\end{table}

The analysis of momenta in the magnetic field allows comparing the ratio of contributions of stable and unstable Be and B isotopes to the fragmentation $^{16}$O + $p$ under identical observational conditions. In Fig. 5, distributions of these fragments are shown with respect to the $P_{\textrm{Be(B)}}$/$P_0$ ratio, which is an estimate of the mass number in the fragmentation cone. For convenience, the data on the decays of $^{8}$Be and $^{9}$B are given with the decreasing (0.5) and increasing (3) factors. Parameterization by Gaussians makes it possible to isolate peaks with half-widths approximately equal to 0.5 and to estimate the isotope statistics. Superimposition of total momentum distributions of 2$\alpha$ pairs $P_{2\alpha}$/$P_0$ for $Q_{2\alpha}$($^{8}$Be) $\leq$ 0.2 MeV and 2$\alpha p$ triplets $P_{2\alpha p}$/$P_0$ for $Q_{2\alpha p}$($^{9}$B) $\leq$ 0.5 MeV displays them in the ranges corresponding to $^{8}$Be and $^{9}$B. Then the statistics for $^{7}$Be, $^{8}$Be, $^{9}$Be, and $^{10}$Be are 196, 345, 92, and 46, and the statistics for $^{8}$B, $^{9}$B, $^{10}$B, $^{11}$B, and $^{12}$B are 33, 60, 226, 257, and 70, respectively. Since the data are obtained uniformly, these values can be used for comparison. The ratio for $^{9}$B and $^{9}$Be mirror nuclei is 0.7 $\pm$ 0.1. However, it is not 1, which indicates differences in the formation of these fragments. At the same time, the fact that the statistics are of the same order of magnitude is an independent argument in favor of correctness of the identification of $^{9}$B in the assumed approximation. 

\section{CORRELATION WITH THE MULTIPLICITY OF $\alpha$ PARTICLES}
Scanning along primary tracks in the NTE makes it possible to detect interactions without sampling, in particular, with different numbers of relativistic He and H fragments. Although the available statistics of multiple channels is many times smaller than that
obtained by the cross scanning, this allows us to trace its evolution with $n_\alpha$ and to assign a reference point for including the results of the accelerated search in the overall picture. Next, measurements based on detecting the tracks of relativistic $^{16}$O, $^{22}$Ne, $^{28}$Si, and $^{197}$Au nuclei in the NTE are used. These data were obtained
by the emulsion collaboration at the JINR synchrophasotron in the 1980s and the EMU collaboration at the AGS (BNL) and SPS (CERN) synchrotrons in the 1990s \cite{7,8,9,10,11}. Photo and video records of characteristic interactions are available \cite{1,12}. The irradiation of NTE with heavier nuclei allows us to extend the multiplicity of relativistic $\alpha$ fragments $n_\alpha$ in the studied events. Data on the ratio of the number of events $N_{n\alpha}$($^{8}$Be) involving at least one $^{8}$Be decay candidate to the $N_{n\alpha}$ channel statistics as a function of $n_\alpha$ are presented in Fig. 5.

There are measurements of interactions of $^{16}$O nuclei at 3.65, 14.6, 60, and 200 GeV/ nucleon. For all values of the initial energy, peak $Q_{2\alpha}$($^{8}$Be) $\leq$ 0.2 MeV is observed \cite{13}. In the initial energy range covered, the $N_{n\alpha}$ and $N_{n\alpha}$($^{8}$Be) distributions show similarity, which allows us to summarize the statistics. The final ratio $N_{n\alpha}$($^{8}$Be)/$N_{n\alpha}$ (\%) increases from $n_{\alpha}$ = 2 (8 $\pm$ 1) to 3 (23 $\pm$ 3) and 4 (46 $\pm$ 14). Its increase is observed in the cases of ``white'' stars $^{12}$C $\to$ 3$\alpha$ and $^{16}$O $\to$ 4$\alpha$ (Fig. 5). Measurements made in the NTE layers exposed to $^{22}$Ne nuclei with the energy of 3.22 GeV/nucleon and with $^{28}$Si nuclei with the energy of 14.6 GeV/nucleon
extend the range of $n_\alpha$ (Fig. 5). In both cases, no change in the condition $Q_{2\alpha}$($^{8}$Be) $\leq$ 0.2 MeV is required. In these cases, the ratio $N_{n\alpha}$($^{8}$Be)/$N_{n\alpha}$ (\%) continues to grow with multiplicity $n_\alpha$ = 2 (6 $\pm$ 1), 3 (19 $\pm$ 3), 4 (31 $\pm$ 6) for $^{22}$Ne and $n_\alpha$ = 2 (3 $\pm$ 2), 3 (13 $\pm$ 5), 4 (32 $\pm$ 6), 5 (38 $\pm$ 11) for $^{28}$Si.

Similar measurements of interactions of $^{197}$Au nuclei at 10.7 GeV/nucleon also indicate that the ratio of the number of events $N_{n\alpha}$($^{8}$Be) involving at least one
$^{8}$Be decay candidate to the channel statistics $N_{n\alpha}$ increases rapidly to about 0.5 by $n_\alpha$ = 10 (Fig. 6). Because of the degraded resolution, the region of $^{8}$Be
expands, which requires softening the selection condition $Q_{2\alpha}$($^{8}$Be) $\leq$ 0.4 MeV. Channels of $n_\alpha$ $\geq$ 11 are summed in order to reduce errors. When the condition is tightened to $Q_{2\alpha}$($^{8}$Be) $\leq$ 0.2 MeV, accompanied by
the decrease in efficiency, the upward trend for $N_{n\alpha}$($^{8}$Be)/$N_{n\alpha}$ is maintained \cite{13}.

\begin{figure}
	\centerline{\includegraphics*[width=1.0\linewidth]{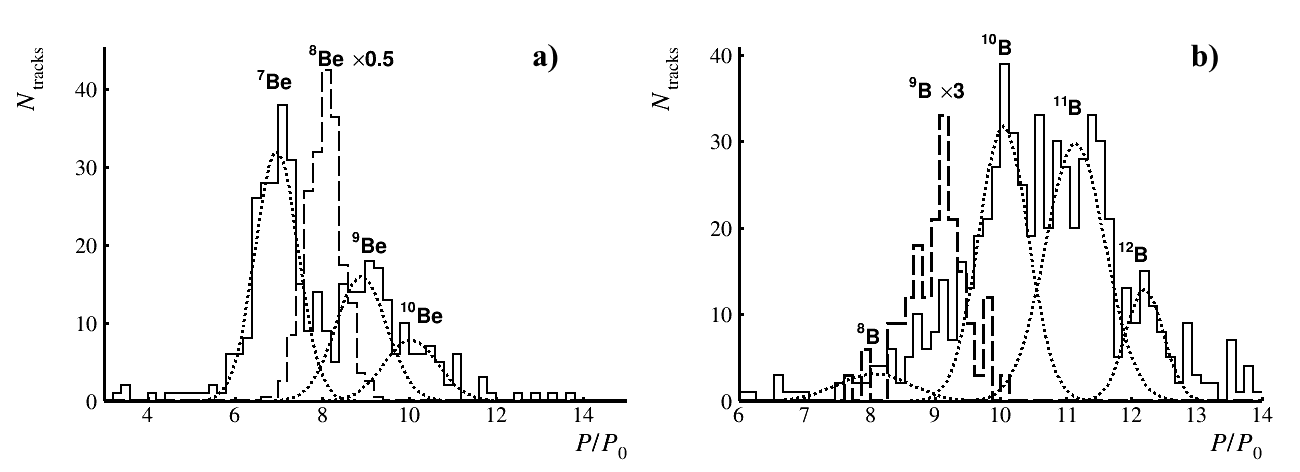}}
	\caption{Distribution of relativistic fragments of (a) Be and (b) B over the ratios of their measured momenta $P_\textrm{H}$ and $P_\textrm{He}$ to the
		initial momentum per nucleon $P_0$ (solid line) in fragmentation of $^{16}$O nuclei with momentum 3.25 GeV/$c$ per nucleon on protons;
		dots show the approximations by the sums of Gaussians; data on $^{8}$Be and $^{9}$B decays are superimposed by dashed lines.}
\end{figure}

\begin{figure}
	\centerline{\includegraphics*[width=1.0\linewidth]{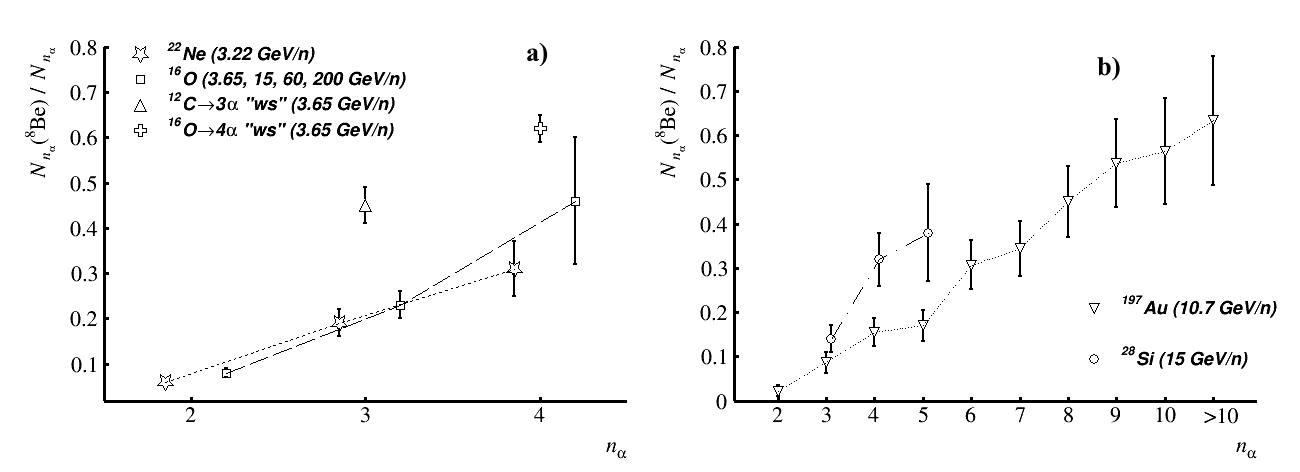}}
	\caption{Dependence of the relative contribution of $N_{n\alpha}$($^{8}$Be) decays to the statistics of $N_n\alpha$ events on the $\alpha$-particle multiplicity $n_\alpha$
		in the relativistic fragmentation of C, O, and Ne nuclei (a) and Si and Au nuclei (b); ``white'' stars $^{12}$C $\to$ 3$\alpha$ and $^{16}$O $\to$ 4$\alpha$ (WS)
		are marked; for convenience, the points are slightly shifted from the values of $n_\alpha$ and connected by dashed lines.}
\end{figure}

The statistics for $^{197}$Au contains 2$\alpha p$ and 3$\alpha$ triplets satisfying the conditions $Q_{2\alpha p}$($^{9}$B) $\leq$ 0.5 MeV and $Q_{3\alpha}$(HS) $\leq$ 0.7 MeV. The ratio of the number of events $N_{n\alpha}$($^{9}$B) and $N_{n\alpha}$(HS) to $N_{n\alpha}$($^{8}$Be) shows no significant change with the variation of multiplicity $n_\alpha$. The statistics of the identified decays of pairs of $^{8}$Be nuclei $N_{n\alpha}$(2$^{8}$Be) behaves in the same way. In fact, these three ratios indicate an increase in $N_{n\alpha}$($^{9}$B), $N_{n\alpha}$(HS), and $N_{n\alpha}$(2$^{8}$Be) with respect to $N_{n\alpha}$. In these three cases, significant statistical errors allow only general trends to be characterized. Summing the statistics of $N_{n\alpha}$($^{9}$B), $N_{n\alpha}$(HS), and $N_{n\alpha}$(2$^{8}$Be) over the multiplicity $n_\alpha$ and
normalizing $N_{n\alpha}$($^{8}$Be) to the sum leads to relative contributions of 25 $\pm$ 4\%, 6 $\pm$ 2\%, and 10 $\pm$ 2\%, respectively.

\section{CONCLUSIONS}
The presented analysis of relativistic $^{16}$O, $^{22}$Ne, $^{28}$Si, and $^{197}$Au nuclei in the nuclear track emulsion indicates that the contribution of the unstable $^{8}$Be nucleus increases with the increase in the number of relativistic $\alpha$ particles. The contributions of the unstable $^{9}$B nucleus decay and the Hoyle state are proportional to that of $^{8}$Be and actually increase. These observations require considering the interactions of $\alpha$ particles generated in the relativistic fragmentation of nuclei. They indicate the intriguing possibility of unstable state fusion reactions between $\alpha$ particles inside relativistic nuclear fragmentation jets. In the
case of the $^{197}$Au nucleus, the upward trend is traced to 10 relativistic $\alpha$ particles per event. In this regard, the statistics of events with an even higher multiplicity of $\alpha$ particles should be increased with the best possible accuracy of measurements of the fragment emission angles.

The results of the study of the $^{16}$O fragmentation in a hydrogen bubble chamber using magnetic analysis confirm the approximations adopted within the invariant mass method. These findings allow a uniform comparison of contributions of stable and unstable nuclei and are suitable for carrying out more complete verification of fragmentation models.

%
%


\begin{thebibliography}{}
%

\bibitem {1}
P. I. Zarubin, Lect. Notes Phys. \textbf{875}, 51 (2013).

\href{https://doi.org/10.1007/978-3-319-01077-9_3}{https://doi.org/10.1007/978-3-319-01077-9\_3}

\bibitem {2}
D. A. Artemenkov et al., Eur. Phys. J. A \textbf{56}, 250 (2020).

\href{https://doi.org/10.1140/epja/s10050-020-00252-3}{https://doi.org/10.1140/epja/s10050-020-00252-3}

\bibitem {3}
A. Tohsaki, H. Horiuchi, P. Schuck, and G. Röpke, Rev. Mod. Phys. \textbf{89}, 011002 (2017).
\href{https://doi.org/10.1103/RevModPhys.89.011002}{https://doi.org/10.1103/RevModPhys.89.011002}

\bibitem {4}
D. A. Artemenkov et al., Rad. Meas. \textbf{119}, 199 (2018).

\href{https://doi.org/10.1016/j.radmeas.2018.11.005}{https://doi.org/10.1016/j.radmeas.2018.11.005}

\bibitem {5}
D. A. Artemenkov et al., Springer Proc. Phys. \textbf{238}, 137 (2020).

\href{https://doi.org/10.1007/978-3-030-32357-8_24}{https://doi.org/10.1007/978-3-030-32357-8\_24}

\bibitem {6}
V. V. Glagolev et al., Eur. Phys. J. A \textbf{11}, 285 (2001).

\href{https://doi.org/10.1007/s100500170067}{https://doi.org/10.1007/s100500170067}

\bibitem {7}
N. P. Andreeva et al., Sov. J. Nucl. Phys. \textbf{47}, 102 (1988).

\bibitem {8}
A. El-Naghy et al., J. Phys. G \textbf(14), 1125 (1988).

\bibitem {9}
M. I. Adamovich et al., Phys. Rev. C \textbf{40}, 66 (1989).

\bibitem {10}
M. I. Adamovich et al., Z. Phys. A \textbf{351}, 311 (1995).

\href{https://doi.org/10.1007/BF01290914}{https://doi.org/10.1007/BF01290914}

\bibitem {11}
M. I. Adamovich et al., Eur. Phys. J. A \textbf{5}, 429 (1999).

\href{https://doi.org/10.1007/s100500050306}{https://doi.org/10.1007/s100500050306}

\bibitem {12}
The BECQUEREL Project.
\href{http://becquerel.jinr.ru/movies/movies.html}{http://becquerel.jinr.ru/movies/movies.html}

\bibitem {13}
A.A. Zaitsev et al., Physics Letters B, \textbf{820}, 136460 (2021)

\href{https://doi.org/10.1016/j.physletb.2021.136460}{https://doi.org/10.1016/j.physletb.2021.136460}



\end{thebibliography}
\end{document}